\documentclass[conference]{IEEEtran}
\usepackage{color}
\usepackage{mathrsfs}
\usepackage{pifont}
\usepackage{bbding}
\usepackage{tipa}
\usepackage{cite}
\usepackage{epsfig}
\usepackage{url}
\usepackage{amsfonts}
\usepackage{multicol}
\usepackage{float}
\usepackage{times}
\usepackage{cite}
\usepackage{psfrag}
\usepackage{subfigure}
\usepackage{stfloats}
\usepackage{amsmath}
\usepackage{footnote}
\usepackage{stmaryrd}
\usepackage{amssymb}
\usepackage{epic}
\usepackage{graphicx}
\usepackage{curves}
\usepackage{algorithm}
\usepackage{algorithmic}
\usepackage{balance}
\usepackage{extarrows}
\usepackage{mathtools}

\newcommand{\Hnull}{\mathcal{H}_0}
\newcommand{\Hone}{\mathcal{H}_1}

\newcommand{\Dnull}{\mathcal{{D}}_0}
\newcommand{\Done}{\mathcal{{D}}_1}

\newtheorem{corollary}{\textbf{Corollary}}
\newtheorem{theorem}{\textbf{Theorem}}

\DeclarePairedDelimiter{\ceil}{\lceil}{\rceil}
\DeclarePairedDelimiter{\floor}{\lfloor}{\rfloor}

\usepackage{cite,graphicx,amsmath,amssymb,cite,algorithm}

\begin{document}


\title{Pilot-Based Channel Estimation Design in Covert Wireless Communication}
\author{
\IEEEauthorblockN{Tingzhen Xu$^{\ast}$, Linlin Sun$^\ast$, Shihao Yan$^{\dag}$, Jinsong Hu$^{\ddag}$, Feng Shu$^{\ast}$}
\IEEEauthorblockA{$^{\ast}$School of Electronic and Optical Engineering, Nanjing University of Science and Technology, Nanjing, China}
\IEEEauthorblockA{$^{\dag}$School of Engineering, Macquarie University, Sydney, Australia}
\IEEEauthorblockA{$^{\ddag}$College of Physics and Information, Fuzhou University,
Fuzhou, China}
\IEEEauthorblockA{Emails: \{tingzhen.xu, sunlinlin, shufeng\}@njust.edu.cn, shihao.yan@mq.edu.au, jinsong.hu@fzu.edu.cn}
}

\vspace{-2cm}

\maketitle
\begin{abstract}
In this work, for the first time, we tackle channel estimation design with pilots in the context of covert wireless communication. Specifically, we consider Rayleigh fading for the communication channel from a transmitter to a receiver and additive white Gaussian noise (AWGN) for the detection channel from the transmitter to a warden. Before transmitting information signals, the transmitter has to send pilots to enable channel estimation at the receiver. Using a lower bound on the detection error probability, we first prove that transmitting pilot and information signals with equal power can minimize the detection performance at the warden, which is confirmed by the minimum detection error probability achieved by the optimal detector based on likelihood ratio test. This motivates us to consider the equal transmit power in the channel estimation and then optimize channel use allocation between pilot and information signals in covert wireless communication. Our analysis shows that the optimal number of the channel uses allocated to pilots increases as the covertness constraint becomes tighter. In addition, our examination shows that the optimal percentage of all the available channel uses allocated to channel estimation decreases as the total number of channel uses increases.
\end{abstract}

\section{Introduction}
Nowadays, the public is increasingly relying on wireless communication for information exchange, which makes the security and privacy of wireless communication being of growing importance. Most techniques in physical layer, which address the security and privacy of wireless communication, focus on protecting the content of a message against eavesdroppers~\cite{bloch2011physical,yan2016artificial}. However, in some special application scenarios, hiding the very existence of wireless transmission or the location information of a transmitter is also critical.
For example, in a military communication scenario, the exposure of the commander's location information is very dangerous~\cite{Sobers2017Covert}, which may lead to fatal attacks. Fortunately, an emerging technique is indeed addressing this problem, which is named as covert communication~\cite{bash2015hiding,yan2019lpdc}. Covert communication aims at guaranteeing that a transmitter (Alice) can send information to a legitimate receiver (Bob) reliably and covertly under the supervision of a warden (Willie), who is detecting whether there exists a wireless transmission or not.

Although covert communication technique is still in its infancy, a couple of aspects of this technique have been studied. For example, the fundamental limits of covert communication was established in \cite{bash2013limits}, which led to a square root law in the context of covert communication. The impact of a full-duplex receiver on covert communication was examined in the literature (e.g., \cite{Hu2017ICC,Khurram2018fullduplex,shu2019delay}), where it was shown that the artificial noise (AN) transmitted by the full-duplex receiver can significantly enhance the performance of covert communication. Most existing works on covert communication focused on point-to-point communication, which ignored the scenario where the distance between Alice and Bob is too large for direct communication. In order to tackle the issues associated with long-distance covert communication, \cite{Azadeh2018Multi} extends the point-to-point covert communication to multi-hop covert communication, aiming at identifying the optimal paths that maximize the covert throughput, and minimizing the end-to-end delay. Meanwhile, covert communication in relay networks were considered in \cite{Hu2018covertrelay} and \cite{Hu2019selfsustained}, where the possibilities and conditions for conducting covert communication in relay networks were determined.

Most existing works on covert communication assumed an infinite blocklength for communication, i.e., the number of channels uses $n$ tends to infinity. However, in most realistic scenarios, the number of channel uses is always finite~\cite{sun2018short-packet}. In these networks, timeless is critical and thus low latency (i.e., short delay) is required. Against this background, \cite{Shihao2018Delay} and \cite{ShihaoYan2017Covert} studied delay-intolerant covert communication in additive white Gaussian noise (AWGN) channel. It proved  that the optimal number of channel uses, which maximizes the covert throughput, should be the maximum allowable number of channel uses $N$. In addition to the aforementioned aspects of covert communication, it was also studied by considering other key affecting factors, including but not limited to, multi-antenna technique~\cite{zheng2019multi}, noise uncertainty~\cite{BiaoHe2017on,goeckel2016covert}, external jammers~\cite{Sobers2017Covert,Biao2018Poisson}, Gaussian signalling strategies~\cite{yan2018gaussian}, and unmanned aerial vehicles~\cite{Zhou2018Joint}. However, channel estimation design has never been considered in the context of covert wireless communication, although it is agreed that how to obtain channel state information (CSI) in covert wireless communication is a challenging research problem. Channel inversion power control was considered in the context of covert wireless communication, which eliminated channel estimation at the receiver based on the pilots transmitted by the transmitter~\cite{Hu2018Inversion}. However, we note that this channel inversion power control requires channel reciprocity; otherwise it cannot provide reliable communication.

In this work, we consider the traditional channel estimation in the context of covert wireless communication, where Alice sends pilot signals before its information transmission in order to enable Bob to estimate CSI. We note that in the traditional channel estimation without any covertness constraint, the transmit power of pilot signals may be different from that of the information signals and the optimal number of channel uses allocated to channel estimation is the same as the number of transmit antennas~\cite{Hassibi2003HowMuch}. However, this may not be true in covert wireless communication, since different transmit power of pilot and information signals may lead to a high detection performance at Willie, who is observing all the transmission (including pilot and information signals) in order to detect this transmission. To examine this issue, we first prove that equal transmit power for pilot and information signals can minimize Willie's detection performance. Motivated by this, we consider equal transmit power for both the pilot and information signals in the channel estimation of covert wireless communication, which results in that we only have to optimize the channel use allocation in the channel estimation design problem. To solve this problem, we analytically derive the optimal number of channel uses allocated channel estimation, which maximizes the effective signal-to-noise ratio (SNR) of the communication channel from Alice to Bob, subject to the covertness constraint. Our examination shows that this optimal number increases as the covertness constraint becomes tighter.


\emph{Notations:} Scalar variables are denoted by italic symbols. Vectors are denoted by lower-case boldface symbols. $\mathbb{E}[\cdot]$ denotes expectation operation. $\mathbf{x}[i]$ denotes the $i$-th element of a vector $\mathbf{x}$.

\section{System Model}

\subsection{Communication Scenario and Adopted Assumptions}

A covert wireless communication system is considered in this work, where Alice tries to send information to Bob, while Willie is monitoring the communication environment to make a decision on whether Alice is transmitting signals to Bob or not.
We consider fading wireless channels from Alice to Bob, where the channel coefficients remain constant in one slot, changing independently from one slot to another, i.e., the channel from Alice to Bob is subject to quasi-static Rayleigh fading.
As such, it is necessary to estimate the channel if Bob conducts coherent decoding. We assume that Alice transmits totally $n$ symbols to Bob in one slot, $n_p$ of which are pilots used to estimate the channel from Alice to Bob for coherent communication and the remaining $n_d = n-n_p$ symbols are data symbols. We also assume that both the pilot sequence and information codebook are kept secret from Willie, i.e., Willie only knows the statistical distribution of them but does not know the realizations of them. The channel between any two users $i$ and $j$ is represented by $h_{ij}$. The subscript $ij$ can be $ab$ or $aw$ corresponding to the Alice-Bob and Alice-Willie channels, respectively.
The AWGN at Bob and Willie are denoted as $n_b[i]$ and $n_w[i]$, respectively, i.e., $\mathbf{n}_b[i]\sim \mathcal{CN}(0,\sigma_b^2)$, $\mathbf{n}_w[i]\sim \mathcal{CN}(0,\sigma_w^2)$, where $\sigma_b^2$ and $\sigma_w^2$ are the noise variances at Bob and Willie, respectively, while $i = 1, 2, ..., n$ is the index of each symbol.
Specifically, we consider that $h_{ab}\sim \mathcal{CN}(0, \lambda_{ab})$ is a zero-mean circularly symmetric complex Gaussian random variable with variance $\mathbb{E}[|h_{ab}|^2]=\lambda_{ab}$. We further assume that the channel between Alice and Willie is AWGN channel, which is motivated by the worst-case scenario for covert wireless communication, where the detection of the transmission from Alice to Bob by Willie becomes the easiest.

\subsection{Binary Detection Problem at Willie}

Willie does not know whether Alice transmits signals to Bob and thus his received signal can be represented as
\begin{equation} \label{yw}
{{\mathbf{y}}_{w}} =\left\{
\begin{aligned}
& {\mathbf{n}_w}, &\Hnull,\\
&\mathbf{x}+\mathbf{n}_w, &\Hone,
\end{aligned}
\right.
\end{equation}
where the null hypothesis $\Hnull$ indicates that Alice does not transmit, the alternative hypothesis $\Hone$ indicates that Alice transmits covert information to Bob, and $\mathbf{x}$ denotes the transmitted signal vector by Alice.
Willie is supervising this transmission action and attempts to determine whether Alice is communicating with Bob or not. Therefore, he has to make a binary decision $\mathcal{D}_0$ or $\mathcal{D}_1$, where $\mathcal{D}_0$ denotes the decision that the received signal is AWGN only and $\mathcal{D}_1$ denotes the decision that the received signal is the signal sent by Alice plus AWGN.
Willie's decision of $\mathcal{D}_1$ on $\Hnull$ causes false alarm and the false alarm rate is denoted by $\alpha$,  while his decision of $\mathcal{D}_0$ on $\Hone$ causes missed detection and the miss detection rate is denoted by $\beta$.
Mathematically, the false alarm rate at Willie is given by
\begin{align} \label{Pfa_Q}
\alpha\triangleq Pr\{\Done|\Hnull\},
\end{align}
and the miss detection rate at Willie is defined as
\begin{align} \label{Pmd_Q}
\beta  \triangleq Pr\{\Dnull|\Hone\}.
\end{align}
For Willie, he attempts to find the optimal detector that minimizes the detection error probability $\xi=\alpha+\beta$, such that he can make a decision on whether Alice is communicating with Bob or not with a high accuracy and the corresponding minimum value of $\xi$ is denoted by $\xi^\ast$. Therefore, the covertness constraint in covert communication is normally written as $\xi^\ast \geq 1 - \epsilon$, where $\epsilon$ is an arbitrarily small value to determine the level of the required covertness.

\subsection{Communication from Alice to Bob}

When Alice transmits signals, the receive signal vector at Bob is given by
\begin{equation} \label{yb}
{\mathbf{y}_{b}} =h_{ab}\mathbf{x}+\mathbf{n}_b.
\end{equation}
When Bob receives signals, he will first estimate the channel with $n_p$ pilots and then decode the data. In this work, we consider the minimum mean square error (MMSE) estimator at Bob for channel estimation. As such, the actual fading coefficient $h_{ab}$ can be represented as the sum of the estimated channel and the estimation error, which are denoted by $\widehat{h}_{ab}$ and $\widetilde{h}_{ab}$, respectively. Therefore, we have\cite{Gursoy2009Onthe}.
\begin{equation} \label{haw_estimate}
{h}_{ab} =\widehat{h}_{ab}+\widetilde{h}_{ab},
\end{equation}
where
\begin{equation} \label{hwidehat}
\widehat{h}_{ab}\sim \mathcal{CN}\left(0,\frac{\lambda_{ab}^2n_p\rho_p}{\lambda_{ab}n_p\rho_p+\sigma_b^2}\right),
\end{equation}
\begin{equation} \label{hwidetilde}
\widetilde{h}_{ab}\sim \mathcal{CN}\left(0,\frac{\lambda_{ab}\sigma_b^2}{\lambda_{ab}n_p\rho_p+\sigma_b^2}\right),
\end{equation}
 and $\rho_p$ represents the average power of the pilot symbols. For convenience, we denote the variance of estimated channel $\widehat{h}_{ab}$ as $\sigma_{\widehat{h}_{ab}}^2$ and denote the variance of estimation error $\widetilde{h}_{ab}$ as $\sigma_{\widetilde{h}_{ab}}^2$.

In the decoding phase, the signal vector at Bob can be rewritten as
\begin{equation}
{\mathbf{y}_{bd}} =\sqrt{\rho_d}\widehat{h}_{ab}\mathbf{x}_d+
\underbrace{\sqrt{\rho_d}\widetilde{h}_{ab}\mathbf{x}_d+\mathbf{n}_b}_{\mathbf{n}_b'},
\end{equation}
where $\rho_d$ is the average power of the data symbols, $\mathbf{x}_d\in \mathcal{C}^{1\times n_d}$ represents the data symbols satisfying $\mathbb{E}\left[\mathbf{x}_d[i]\mathbf{x}_d^H[i]\right]=1$, $i=1,2,...,n-n_p$, and $\mathbf{n}_b'\in \mathcal{C}^{1\times n_d}$ presents the effective noise for decoding. Apparently, the variance of $\mathbf{n}_b'$ is given by
\begin{equation} \label{noise_var}
\sigma_{\mathbf{n}_b'}^2=\frac{1}{n_d}\mathbb{E}\left[\mathbf{n}_b^H\mathbf{n}_b'\right]
=\sigma_b^2+\rho_d\sigma_{\widetilde{h}_{ab}}^2.
\end{equation}
Then, following \eqref{hwidehat} and \eqref{noise_var},  the signal-to-interference-plus-noise ratio (SINR) can be written as
\begin{align} \label{rho_eff}
\gamma \triangleq&\frac{\rho_d\sigma_{\widehat{h}_{ab}}^2}
{\sigma_b^2+\rho_d\sigma_{\widetilde{h}_{ab}}^2}
=&\frac{\rho_d\lambda_{ab}^2n_p\rho_p}{\sigma_b^4+\rho_d\lambda_{ab}\sigma_b^2
+\sigma_b^2\lambda_{ab}n_p\rho_p}.
\end{align}
To simplify the presentation, we denote the power fraction allocated to the data from the total power as $\eta$, and $1-\eta$ denotes the fraction of the total power that is allocated to pilot symbols\cite{Hassibi2003HowMuch}, i.e., we have
\begin{equation} \label{alpha}
\rho_dn_d=\eta\rho n, ~\rho_pn_p=(1-\eta)\rho n, ~0<\eta<1,
\end{equation}
where we recall that $n_p+n_d=n$ and $\rho$ is the average power of all the transmitted symbols. Considering the cost of channel estimation, the effective SINR per channel use can be written as
\begin{equation} \label{capacity}
\gamma_{\text{eff}} =\frac{n-n_p}{n}\times \gamma,
\end{equation}
where, as per \eqref{rho_eff} and \eqref{alpha}, $\gamma$ can be rewritten as
\begin{equation} \label{rho_eff_alpha}
\gamma =\frac{\lambda_{ab}^2\rho n \eta (1-\eta)}{(n-n_p)\sigma_b^2\left[\lambda_{ab}(1-\eta)+\frac{\sigma_b^2}
{\rho n}+\frac{\lambda_{ab}\eta}{n-n_p}\right]}.
\end{equation}

\section{Optimal Channel Use Allocation for Covert Wireless Communication}

In this section, we tackle the optimal resource allocation in the context of covert wireless communication with channel estimation. Specifically, we first analyze the detection performance at Willie, which indicates that the equal transmit power for pilot and information signals can minimize Willie's detection performance. This motivates us to optimize the number of channel uses allocated between the pilot and information symbols with equal transmit power.

\subsection{Detection Performance at Willie}

Under $\Hnull$, the receive signal vector at Willie consists of AWGN only and thus the likelihood function of $\mathbf{y}_w$ is given by
\begin{align}
f(\mathbf{y}_w|\Hnull)=\frac{1}{(2\pi\sigma_w^2)^{n/2}}\exp\left({-\frac{ \sum_{i=1}^{n}{y}_w^i}
{2\sigma_w^2}}\right),
\end{align}
where ${y}_w^i$ denotes the $i$-th element of $\mathbf{y}_w$. Under $\Hone$, the receive signal vector at Willie can be specifically written as
\begin{equation}
\mathbf{y}_w=[y_w^1,...,y_w^{n_p},\underbrace{y_w^{n_p+1}...,y_w^{n}}_{n_d}],
\end{equation}
where the first $n_p$ symbols consist of Alice's transmitted pilots and AWGN and the following $n_d$ symbols consist of Alice's transmitted information signals and AWGN. As such, the likelihood function of $\mathbf{y}_w$ under $\Hone$ is given by
\begin{align} \label{f_likeli}
f(\mathbf{y}_w|\Hone)=&\prod_{i=1}^{n_p}f({y}_w^i)\notag
\prod_{i=n_p+1}^{n}f(y_w^i)\\ \nonumber
=&\frac{1}{(2\pi(\rho_p+\sigma_w^2))^{n_p/2}}e^{-\frac{\sum_{i=1}^{n_p}{y}_w^i}
{2(\rho_p+\sigma_w^2)}} \\
&\times \frac{1}{(2\pi(\rho_d+\sigma_w^2))^{\frac{n-n_p}{2}}}
e^{\frac{{-\sum_{i=n_p+1}^{n}{y}_w^i}}{{2(\rho_d+\sigma_w^2)}}}.
\end{align}

Due to the high complexity of \eqref{f_likeli}, it is hard to derive the miss detection rate in a closed-form expression, which leads to the fact that the covertness constraint, i.e., $\xi^\ast \geq 1 - \epsilon$, cannot be explicitly determined. Fortunately, this covertness constraint can be guaranteed by $\mathcal{D}_{01} \leq 2\epsilon^2$ due to $\xi^\ast \geq 1 - \sqrt{\mathcal{D}_{01}/2}$ \cite{bash2013limits,Shihao2018Delay,shu2019delay}, where $\mathcal{D}_{01}$ is the KL divergence from $f(\mathbf{y}_w|\Hnull)$ to $f(\mathbf{y}_w|\Hone)$, which is given by \cite{Shihao2018Delay,shu2019delay}
\begin{align}\label{d01_definition}
&\mathcal{D}_{01}=n_p\left[\ln{(1\!+\!\frac{\rho_p}{\sigma_w^2})}\!-\!\frac{\frac{\rho_p}{\sigma_w^2}}{1\!+\!\frac{\rho_p}{\sigma_w^2}}\right]\!+\!
n_d\left[\ln{(1\!+\!\frac{\rho_d}{\sigma_w^2})}\!-\!\frac{\frac{\rho_d}{\sigma_w^2}}{1\!+\!\frac{\rho_d}{\sigma_w^2}}\right]\notag\\
& \xlongequal {a}\frac{(1-\eta)n\rho n_p}{(\eta-1)n\rho\!-\!n_p\sigma_w^2}+n_d\left[\ln{(1\!+\!\frac{n\rho\eta}{n_d\sigma_w^2})
\!-\!\frac{n\rho\eta}{n\rho\eta+n_d\sigma_w^2}}\right] \notag\\
& ~~~~+n_p\ln{\left[1+\frac{n\rho(1-\eta)}{n_p\sigma_w^2}\right]},
\end{align}
and $\xlongequal {a}$ is obtained following \eqref{alpha} and conducting some simplifications. In the following, we show that $\mathcal{D}_{01}$ is minimized when $\rho_p = \rho_d$, i.e., when $\eta = n_d/n$. To this end, we first derive the first derivative of $\mathcal{D}_{01}$ with respect to $\eta$ as
\begin{align}\label{d01_eta}
\frac{\partial \mathcal{D}_{01}}{\partial \eta}=\frac{n^2(n\eta-n_d)(n^2\rho^2\eta^2-n^2\rho^2\eta+n_dn_p\sigma_w^4)}
{(n\eta\rho+n_d\sigma_w^2)^2[(1-\eta)n\rho+n_p\sigma_w^2]^2}.
\end{align}
Following \eqref{d01_eta}, we can see that $\eta={n_d}/{n}$, which leads to $\rho_p = \rho_d$ as per \eqref{alpha}, is one reasonable solution to ${\partial \mathcal{D}_{01}}/{\partial \eta} =0$, since other solutions require $\rho \geq \sigma_w^2$ that leads to a very low detection error probability (i.e., a low value of $\xi^\ast$) at Willie, which is not feasible in the context of covert communication. This result indicates that the equal transmit power of pilot and information signals can be potentially the optimal power allocation strategy in the channel estimation for covert wireless communication, since the covertness constraint is normally the main performance limiting factor in covert communication and thus in the design of covert communication system the first priority is to minimize the detection performance at Willie. As such, in this work and the following subsection we consider the equal transmit power and then optimize the channel use allocation to maximize the system performance.

\subsection{Optimal Channel Use Allocation for $\rho_p=\rho_d=\rho$}

As we proved in the last subsection, $\rho_p=\rho_d=\rho$ can minimize the detection performance at Willie. Thus, we consider channel use allocation in this subsection with equal transmit power for pilot and data signals. With $\rho_p=\rho_d=\rho$, a compact expression of Willie's minimum detection error probability $\xi^\ast$ can be derived and thus we use $\xi^\ast \geq 1 - \epsilon$ as the covertness constraint. Then,  the optimization problem at Alice is given by
\begin{align}
\underset{\rho, n_p}{\max} \quad &\gamma_{\text{eff}}, \label{ObjectFuncOne}\\
\text{s. t.} \quad  &\xi^{\ast}\geq1-\epsilon, \\\nonumber
& n_p=1,2,\ldots,n-1. \nonumber
\end{align}
The solution to this optimization problem is given in the following theorem.
\begin{theorem}
\label{theorem1}
With $\rho_p=\rho_d=\rho$, the optimal number of pilot symbols $n_p$ that maximizes the effective SINR $\gamma_{\text{eff}}$ is given by
\begin{equation}
\label{optimal_np}
{n_p^\ast} =\left\{
\begin{aligned}
&n_p^{\mathrm{ceil}}, &\text{if}~\gamma_{\text{eff}}^{\mathrm{ceil}}\geq \gamma_{\text{eff}}^{\mathrm{floor}},\\
&n_p^{\mathrm{floor}},&\text{if}~\gamma_{\text{eff}}^{\mathrm{ceil}}< \gamma_{\text{eff}}^{\mathrm{floor}},
\end{aligned}
\right.
\end{equation}
and the optimal value of the average power $\rho$, i.e., $\rho^\ast$, is the solution to the following equation
\begin{equation}\label{optimal_rho}
\frac{\gamma\left(n,\frac{n(\rho+\sigma_w^2)}{\rho}\ln(1+\frac{\rho}{\sigma_w^2})\right)}
{\Gamma(n)}-\frac{\gamma\left(n,\frac{n\sigma_w^2}{\rho}\ln(1+\frac{\rho}{\sigma_w^2})\right)}
{\Gamma(n)}=\epsilon,
\end{equation}
where
\begin{equation}
\begin{aligned}
&n_p^{\mathrm{ceil}}=\ceil[\bigg]{\frac{-\kappa+\sqrt{\kappa(\kappa+n\lambda_{ab}\rho^\ast)}}{\lambda_{ab}\rho^\ast}},\\
&n_p^{\mathrm{floor}}=\floor[\bigg]{\frac{-\kappa+\sqrt{\kappa(\kappa+n\lambda_{ab}\rho^\ast)}}{\lambda_{ab}\rho^\ast}},
\end{aligned}
\end{equation}
while $\ceil{\cdot}$ denotes the ceiling function,  $\floor{\cdot}$ denotes the floor function, $\gamma_{\text{eff}}^{\mathrm{ceil}}$ and $\gamma_{\text{eff}}^{\mathrm{floor}}$ are obtained by substituting $n_p^{\mathrm{ceil}}$ and $n_p^{\mathrm{floor}}$ into \eqref{capacity} with $\rho_p = \rho_d = \rho^\ast$, respectively, and $\kappa=\lambda_{ab}\rho^\ast+\sigma_b^2$.
\end{theorem}
\begin{IEEEproof}
Under the special case of $\rho_p=\rho_d=\rho$, the test statistic $T$ can be written as
\begin{equation}
T=\frac{\rho+\sigma_w^2}{2n}(\chi_{2n_p}^2+\chi_{2n_d}^2)=
\frac{\rho+\sigma_w^2}{2n}\chi_{2n}^2,
\end{equation}
which becomes a chi square random variable with $2n$ degrees of freedom. The optimal detection threshold can be directly obtained as \cite{Shihao2018Delay}
\begin{align}
\tau^{\ast}=\frac{\sigma_w^2}{\rho}(\sigma_w^2+\rho)\ln(1+\frac{\rho}{\sigma_w^2}).
\end{align}
Then, the minimum detection error probability at Willie is given by~\cite{Shihao2018Delay,Lee2015Achieving}
\begin{align}\label{minimum_xi}
\xi^{\ast}=1-\frac{\gamma(n,\frac{n\tau^{\ast}}
{\sigma_w^2})}{\Gamma(n)}+\frac{\gamma\left(n,\frac{n\tau^{\ast}}
{\rho+\sigma_w^2}\right)}{\Gamma(n)},
\end{align}
where $\Gamma(\cdot)$ is the complete gamma function and $\gamma(\cdot,\cdot)$ is the lower incomplete gamma function. 
Again, noting the equality in the covertness constraint is always guaranteed, the optimal $\rho$ is achieved by solving $\xi^{\ast}=1-\epsilon$, which is shown in \eqref{optimal_rho}.
As per \eqref{Pfa_Q} and \eqref{minimum_xi}, we can see that in this case, $\xi^\ast$ is affected by $n$, neither $n_p$ nor $n_d$. This is due to the fact that when $\rho_p=\rho_d=\rho$, the pilot and data are the same for Willie, since they are unknown by Willie, and only their total number will affect the detection performance of Willie. As such, the optimal value of $\rho$ is not a function of $n_p$ or $n_d$, which leads to the fact that $\rho^\ast$ does not depend on $n_p$ in our considered optimization problem.

We next derive the optimal value of $n_p$. When $\rho_p=\rho_d=\rho^\ast$, the effective SINR given in \eqref{capacity} can be rewritten as
\begin{equation}\label{eff_rho}
\gamma_{\text{eff}}=\frac{\lambda_{ab}^2{\rho^\ast}^2n_p}{\sigma_b^2\left[\sigma_b^2
+\lambda_{ab}\rho^\ast (n_p+1)\right]},
\end{equation}
Then the first derivative of $\gamma_{\text{eff}}$ with respect to $n_p$ is derived as
\begin{align}\label{dC/dnp}
&\frac{\partial \gamma_{\text{eff}}}{\partial n_p}= \notag\\
&\frac{\lambda_{ab}^2{\rho^\ast}^2\left[\!-\!\lambda_{ab}\rho^\ast n_p^2-(2\sigma_b^2+2\lambda_{ab}\rho^\ast)n_p+n\sigma_b^2+\lambda_{ab}\rho^\ast n\right]}{n\ln2\left[\sigma_b^3
+\lambda_{ab}\rho^\ast\sigma_b(n_p+1)\right]^2}.
\end{align}
We observe that the term in the square brackets of \eqref{dC/dnp} is a quadratic equation. It can be easily proved that there are always two real solutions to $\frac{\partial \gamma_{\text{eff}}}{\partial n_p} =0$, one of which is positive and the other is negative. Specifically, due to the fact that $n_p$ is positive, we keep the positive one, which is given by
\begin{equation}\label{np_raw}
n_p=\frac{-\kappa+\sqrt{\kappa(\kappa+n\lambda_{ab}\rho^\ast)}}{\lambda_{ab}\rho^\ast}.
\end{equation}
We note that the number of pilot symbols should be an integer, which leads to the result given in \eqref{optimal_np}.
This completes the proof of Theorem~1.
\end{IEEEproof}

Following Theorem~1, we have the following corollary on the impact of the covertness constraint on the optimal value of $n_p$, i.e., the number of channel uses allocated to channel estimation in covert wireless communication.
\begin{corollary}
As $\epsilon$ decreases, the optimal value of $n_p$ potentially increases (definitely does not decrease), which indicates that more channel uses would be allocated to  channel estimation as the covertness constraint becomes stricter.
\end{corollary}
\begin{IEEEproof}
Considering the ceiling and floor functions, we prove this corollary by proving that the value of $n_p$ given in \eqref{np_raw} monotonically decreases with $\epsilon$. The first derivative of $n_p$ given in \eqref{np_raw} with respect to $\rho^\ast$ is derived as
\begin{equation}\label{coro_p1}
\frac{\partial n_p}{\partial\rho^\ast}=-\frac{\sigma_b^2\left[2\kappa+n\rho^\ast\lambda_{ab}-
2\sqrt{\kappa(\kappa+n\rho^\ast\lambda_{ab})}\right]}
{2\lambda_{ab}{\rho^\ast}^2\sqrt{\kappa(\kappa+n\rho^\ast\lambda_{ab})}}.
\end{equation}
Following \eqref{coro_p1}, we note that $\frac{\partial n_p}{\partial\rho}<0$, due to
\begin{align}
\left[2\kappa+n\rho^\ast\lambda_{ab}\right]^2-
\left[2\sqrt{\kappa(\kappa+n\rho^\ast\lambda_{ab})}\right]^2 = (n\rho^\ast\lambda_{ab})^2 >0,\notag
\end{align}
$2\kappa+n\rho^\ast\lambda_{ab}>0$, and $2\sqrt{\kappa(\kappa+n\rho^\ast\lambda_{ab})}>0$.
Again, due to the fact that equality in the covertness constraint is always guaranteed, i.e., $\xi^\ast = 1 - \epsilon$, $\rho^\ast$ is a monotonically increasing function of $\epsilon$. This completes the proof of Corollary~1.
\end{IEEEproof}

\section{Simulations and Discussions}

In this section, we provide numerical results to verify our analysis and examine the impact of different system parameters on the channel estimation design in the context of covert wireless communication.

\begin{figure}[t]
\centering
\includegraphics[height=2.9in, width=3.6in]{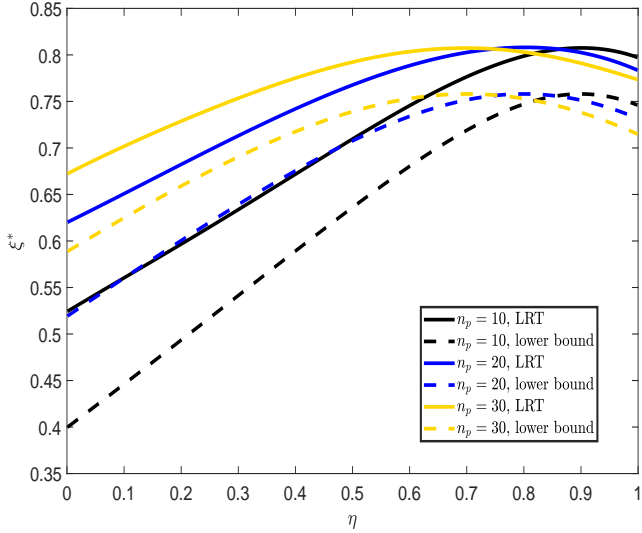}\\
\caption{The minimum detection error probability $\xi^\ast$ and the lower bound determined by the KL divergence $\mathcal{D}_{01}$ versus the power allocation parameter $\eta$ for different values of the number of channel uses allocated to channel estimation (i.e., $n_p$), where $\rho=0.05$, $n=100$, and $\sigma_w^2=0$ dBm.}\label{fig:fig1}
\end{figure}

In order to confirm that equal transmit power for pilot and information signals is optimal in terms of forcing Willie to have the worst detection performance,  in Fig.~\ref{fig:fig1} we plot the minimum detection error probability $\xi^\ast$ achieved by the optimal detector (i.e., the likelihood ratio test) and the lower bound on $\xi^\ast$ determined by the KL divergence given in \eqref{d01_definition}, i.e., $1-\sqrt{\mathcal{D}_{01}/2}$, versus $\eta$ for different values of the number of channel uses allocated to channel estimation (i.e., $n_p$). In this figure, we first observe that $\xi^\ast$ has the same trend as $1-\sqrt{\mathcal{D}_{01}/2}$ with respect to $\eta$, i.e., they are maximized at the same value of $\eta$. We note that, for the values of $\eta$ that maximizes $\xi^\ast$ or $1-\sqrt{\mathcal{D}_{01}/2}$, we have $\rho_p = \rho_d$. This first confirms our analysis based on $\mathcal{D}_{01}$, which shows that  $\rho_p = \rho_d$ minimizes $\mathcal{D}_{01}$. In addition, this figure demonstrates that $\rho_p = \rho_d$ also maximizes the actual minimum detection error probability $\xi^\ast$. As such, we can conclude that equal transmit power indeed minimizes the detection performance at Willie, which motivates us to consider the equal transmit power for pilot and information signals in the channel estimation design for covert wireless communication.

\begin{figure}[t]
\centering
\includegraphics[height=2.9in, width=3.6in]{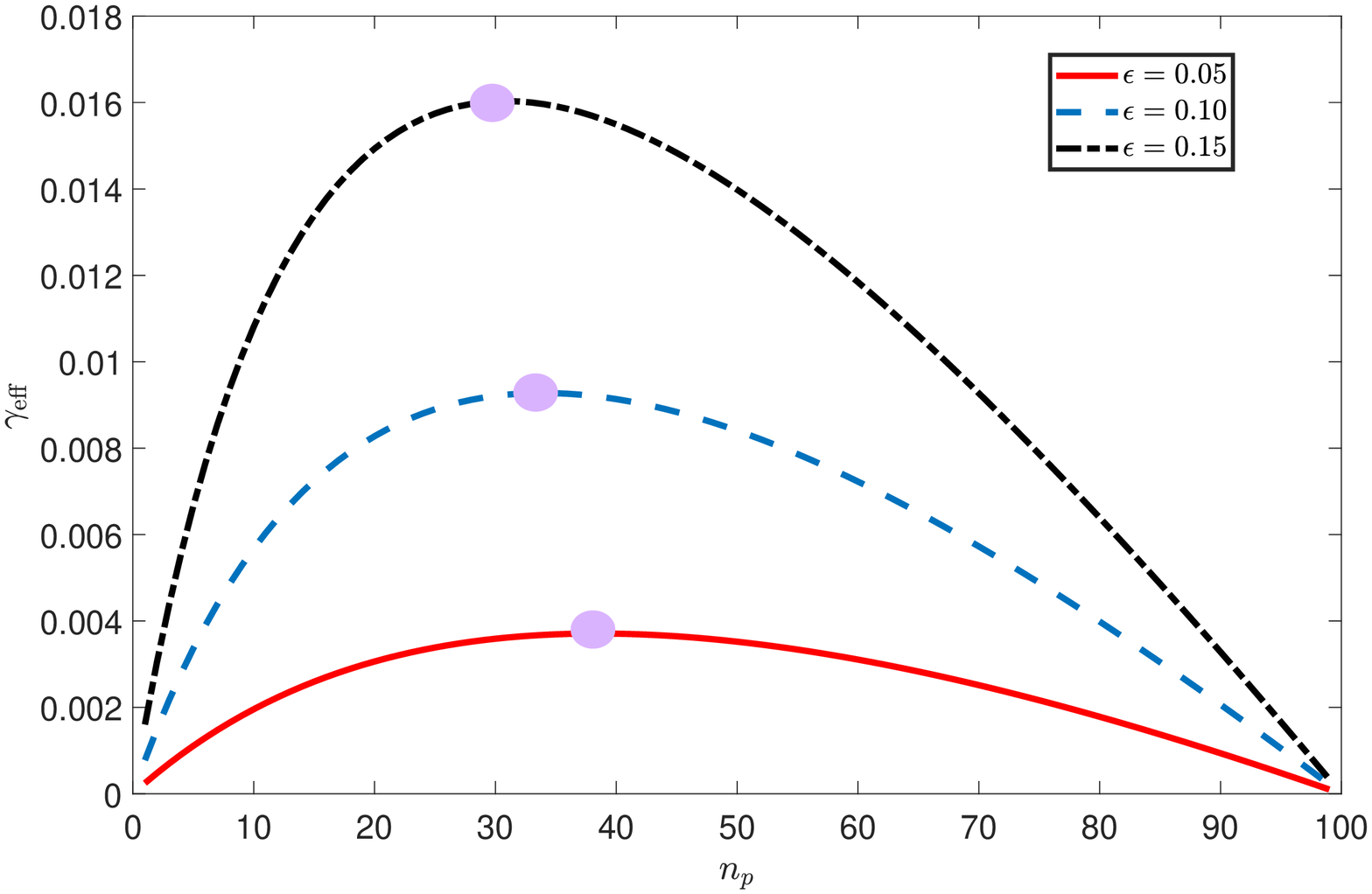}\\
\caption{Effective SINR $\gamma_{\text{eff}}$ versus the number of channel uses for channel estimation (i.e., $n_p$) for different values of the covertness parameter $\epsilon$ with $\rho_p=\rho_d=\rho$, where $n=100$, $\lambda_{ab}=1$, $\tau=0$ dBm, $\sigma_b^2=\sigma_w^2=0$ dBm.}\label{fig:fig2}
\end{figure}

Considering equal transmit power for pilot and information signals, in Fig.~\ref{fig:fig2} we plot the effective SINR $\gamma_{\text{eff}}$ achieved subject to the covertness constraint $\xi^\ast \geq 1 - \epsilon$ versus  $n_p$ for different values of $\epsilon$. In this figure, we first observe that there exists an optimal value of $n_p$ that maximizes $\gamma_{\text{eff}}$ subject to the covertness constraint. We note that in this figure the solid circles denote the analytical optimal values of $n_p$. As such, this figure confirms the correctness of our analysis, i.e., the analytical solutions exactly match with the numerical ones. In addition, in this figure we observe that the maximum $\gamma_{\text{eff}}$ significantly decreases as $\epsilon$ decreases, i.e., as the covertness becomes tighter, which confirms that the key factor that limits the performance of covert communication is the covertness constraint. This is the main reason why we consider the equal transmit power in the channel estimation for covert wireless communication, as the equal transmit power can minimize the detection performance of Willie, i.e., making the covertness constraint being the easiest to satisfy.

\begin{figure}[t]
\centering
\includegraphics[height=2.9in, width = 3.6in]{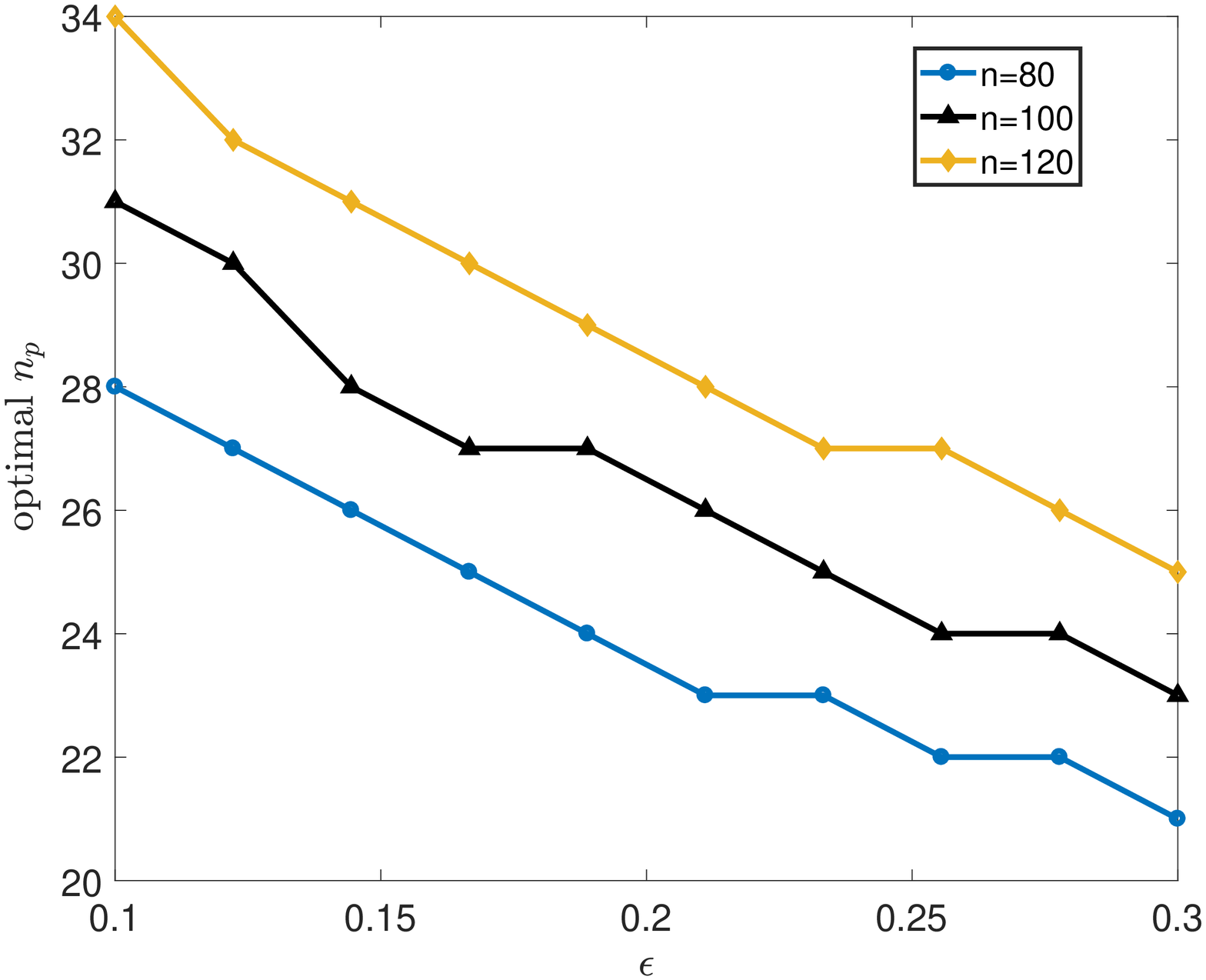}\\
\caption{Optimal number of channel uses allocated to channel estimation (i.e., $n_p^\ast$) versus the covertness parameter $\epsilon$ for different values of the total number of channel uses (i.e., $n$) with $\rho_p=\rho_d=\rho$, where $\lambda_{ab}=1$, $\tau=0$ dBm, and $\sigma_b^2=\sigma_w^2=0$ dBm.}\label{fig:fig3}
\end{figure}

In Fig.~\ref{fig:fig3}, we plot the optimal number of channel uses allocated to channel estimation (i.e., $n_p^\ast$) versus the covertness parameter $\epsilon$ for different values of the total number of channel uses (i.e., $n$) with the equal transmit power. In this figure, we first observe that $n_p^\ast$ monotonically decreases with $\epsilon$, which indicates that $n_p^\ast$ increases as the covertness constraint becomes stricter. Intuitively, this is due to the fact that, as the covertness becomes tighter, the transmit power decreases and we have to increase the number of channel uses to obtain a certain level of channel estimation accuracy. In this figure, as expected we also observe that $n_p^\ast$ increases as $n$ increases. However, we note that the ratio from $n_p^\ast$ to $n$ (i.e., $n_p^\ast/n$) decreases with $n$. This is consistent with the general conclusion that the channel estimation cost in terms of the channel uses is negligible as $n \rightarrow \infty$. We note that this cost cannot be ignored when $n$ is finite and small, especially in the context of covert wireless communication. This is due to the fact that the covertness constraint may enforce equal transmit power for pilot and information signals.


\section{Conclusions}

This work, for the first time, considered the traditional channel estimation in the context of covert wireless communication, where the impact of a finite blocklength was examined. Using a KL divergence to determine a lower bound on the minimum detection error probability at Willie, we proved that equal transmit power for pilot and information signals can minimize Willie's detection performance. This conclusion inspired us to consider the equal transmit power in the channel estimation for covert wireless communication, as the key factor that limits the performance of covert communication is the
covertness constraint. We then derived the optimal number of channel uses allocated to channel estimation in order to maximize the effective SINR of the communication channel subject to the covertness constraint. This optimal number monotonically increases as the covertness constraint becomes tighter, which demonstrates more resource should be allocated to channel estimation in covert wireless communication when more covertness is desired.


\end{document}